\documentclass[prl,nofootinbib,preprint,superscriptaddress]{revtex4}

\usepackage{graphicx}
\usepackage{amsmath}
\usepackage{amsfonts}
\usepackage{amssymb}
\usepackage{dsfont}
\usepackage{dcolumn}
\usepackage{bm}
\usepackage{pstricks}
\usepackage{color}

\newcommand{\beqn}{\begin{eqnarray}}
\newcommand{\eeqn}{\end{eqnarray}}
\newcommand{\be}{\begin{equation}}
\newcommand{\ee}{\end{equation}}

\newcommand{\mathsym}[1]{{}}

\def\br{\left(\begin{array}{c}}
\def\er{\end{array}\right)}

\def\rmuu{\gamma^{\mu}}
\def\rmud{\gamma_{\mu}}
\def\PL{{1-\gamma_5\over 2}}
\def\PR{{1+\gamma_5\over 2}}
\def\sinW2{\sin^2\theta_W}
\def\AEM{\alpha_{EM}}
\def\mul{M_{\tilde{u} L}^2}
\def\mur{M_{\tilde{u} R}^2}
\def\mdl{M_{\tilde{d} L}^2}
\def\mdr{M_{\tilde{d} R}^2}
\def\mz2{M_{z}^2}
\def\c2b{\cos 2\beta}
\def\au{A_u}
\def\ad{A_d}
\def\cob{\cot \beta}
\def\v#1{v_#1}
\def\tb{\tan\beta}
\def\epem{$e^+e^-$}
\def\KK{$K^0$-$\overline{K^0}$}
\def\wi{\omega_i}
\def\xj{\chi_j}
\def\Wmu{W_\mu}
\def\Wnu{W_\nu}
\def\m#1{{\tilde m}_#1}
\def\mH{m_H}
\def\mw#1{{\tilde m}_{\omega #1}}
\def\mx#1{{\tilde m}_{\chi^{0}_#1}}
\def\mc#1{{\tilde m}_{\chi^{+}_#1}}
\def\mwi{{\tilde m}_{\omega i}}
\def\mxi{{\tilde m}_{\chi^{0}_i}}
\def\mci{{\tilde m}_{\chi^{+}_i}}

\def\ch{{\tilde\chi^{+}_1}}
\def\c2{{\tilde\chi^{+}_2}}

\def\tt{{\tilde\theta}}

\def\tp{{\tilde\phi}}

\def\mz{M_z}
\def\sw{\sin\theta_W}
\def\cw{\cos\theta_W}
\def\cb{\cos\beta}
\def\sb{\sin\beta}
\def\rwi{r_{\omega i}}
\def\rxj{r_{\chi j}}
\def\rfp{r_f'}
\def\Kik{K_{ik}}
\def\Fq2{F_{2}(q^2)}
\def\f{\({\cal F}\)}
\def\d1{{\f(\tilde c;\tilde s;\tilde W)+ \f(\tilde c;\tilde \mu;\tilde W)}}
\def\tw{\tan\theta_W}
\def\sec2w{sec^2\theta_W}
\def\lsim{\ ^<\llap{$_\sim$}\ }
\def\gsim{\ ^>\llap{$_\sim$}\ }
\def\r2{\sqrt 2}
\def\beq{\begin{equation}}
\def\eeq{\end{equation}}
\def\beqn{\begin{eqnarray}}
\def\eeqn{\end{eqnarray}}
\def\rmuu{\gamma^{\mu}}
\def\rmud{\gamma_{\mu}}
\def\PL{{1-\gamma_5\over 2}}
\def\PR{{1+\gamma_5\over 2}}
\def\sinW2{\sin^2\theta_W}
\def\AEM{\alpha_{EM}}
\def\mul{M_{\tilde{u} L}^2}
\def\mur{M_{\tilde{u} R}^2}
\def\mdl{M_{\tilde{d} L}^2}
\def\mdr{M_{\tilde{d} R}^2}
\def\mz2{M_{z}^2}
\def\c2b{\cos 2\beta}
\def\au{A_u}         
\def\ad{A_d}
\def\cob{\cot \beta}
\def\v#1{v_#1}
\def\tb{\tan\beta}
\def\epem{$e^+e^-$}
\def\KK{$K^0$-$\bar{K^0}$}
\def\wi{\omega_i}
\def\xj{\chi_j}
\def\Wmu{W_\mu}
\def\Wnu{W_\nu}
\def\m#1{{\tilde m}_#1}
\def\mH{m_H}
\def\mw#1{{\tilde m}_{\omega #1}}
\def\mx#1{{\tilde m}_{\chi^{0}_#1}}
\def\mc#1{{\tilde m}_{\chi^{+}_#1}}
\def\mwi{{\tilde m}_{\omega i}}
\def\mxi{{\tilde m}_{\chi^{0}_i}}
\def\mci{{\tilde m}_{\chi^{+}_i}}
\def\mz{M_z}
\def\sw{\sin\theta_W}
\def\cw{\cos\theta_W}
\def\cb{\cos\beta}
\def\sb{\sin\beta}
\def\rwi{r_{\omega i}}
\def\rxj{r_{\chi j}}
\def\rfp{r_f'}
\def\Kik{K_{ik}}
\def\Fq2{F_{2}(q^2)}
\def\f{\({\cal F}\)}
\def\d1{{\f(\tilde c;\tilde s;\tilde W)+ \f(\tilde c;\tilde \mu;\tilde W)}}
\def\tw{\tan\theta_W}
\def\sec2w{sec^2\theta_W}
\def\ch{{\tilde\chi^{+}_1}}
\def\c2{{\tilde\chi^{+}_2}}

\def\tt{{\tilde\theta}}

\def\tp{{\tilde\phi}}

\def\mz{M_z}
\def\sw{\sin\theta_W}
\def\cw{\cos\theta_W}
\def\cb{\cos\beta}
\def\sb{\sin\beta}
\def\rwi{r_{\omega i}}
\def\rxj{r_{\chi j}}
\def\rfp{r_f'}
\def\Kik{K_{ik}}
\def\Fq2{F_{2}(q^2)}
\def\f{\({\cal F}\)}
\def\d1{{\f(\tilde c;\tilde s;\tilde W)+ \f(\tilde c;\tilde \mu;\tilde W)}}
\def\tw{\tan\theta_W}
\def\sec2w{sec^2\theta_W}

\def\rmuu{\gamma^{\mu}}
\def\rmud{\gamma_{\mu}}
\def\PL{{1-\gamma_5\over 2}}
\def\PR{{1+\gamma_5\over 2}}
\def\sinW2{\sin^2\theta_W}
\def\AEM{\alpha_{EM}}
\def\mul{M_{\tilde{u} L}^2}
\def\mur{M_{\tilde{u} R}^2}
\def\mdl{M_{\tilde{d} L}^2}
\def\mdr{M_{\tilde{d} R}^2}
\def\mz2{M_{z}^2}
\def\c2b{\cos 2\beta}
\def\au{A_u}
\def\ad{A_d}
\def\cob{\cot \beta}
\def\v#1{v_#1}
\def\tb{\tan\beta}
\def\epem{$e^+e^-$}
\def\KK{$K^0$-$\overline{K^0}$}
\def\wi{\omega_i}
\def\xj{\chi_j}
\def\Wmu{W_\mu}
\def\Wnu{W_\nu}
\def\m#1{{\tilde m}_#1}
\def\mH{m_H}
\def\mw#1{{\tilde m}_{\omega #1}}
\def\mx#1{{\tilde m}_{\chi^{0}_#1}}
\def\mc#1{{\tilde m}_{\chi^{+}_#1}}
\def\mwi{{\tilde m}_{\omega i}}
\def\mxi{{\tilde m}_{\chi^{0}_i}}
\def\mci{{\tilde m}_{\chi^{+}_i}}

\def\ch{{\tilde\chi^{+}_1}}
\def\c2{{\tilde\chi^{+}_2}}

\def\tt{{\tilde\theta}}

\def\tp{{\tilde\phi}}

\def\mz{M_z}
\def\sw{\sin\theta_W}
\def\cw{\cos\theta_W}
\def\cb{\cos\beta}
\def\sb{\sin\beta}
\def\rwi{r_{\omega i}}
\def\rxj{r_{\chi j}}
\def\rfp{r_f'}
\def\Kik{K_{ik}}
\def\Fq2{F_{2}(q^2)}
\def\f{\({\cal F}\)}
\def\d1{{\f(\tilde c;\tilde s;\tilde W)+ \f(\tilde c;\tilde \mu;\tilde W)}}
\def\tw{\tan\theta_W}
\def\sec2w{sec^2\theta_W}
\begin{document}
\baselineskip 18pt
\def\today{\ifcase\month\or
 January\or February\or March\or April\or May\or June\or
 July\or August\or September\or October\or November\or December\fi
 \space\number\day, \number\year}
\def\thebibliography#1{\section*{References\markboth
 {References}{References}}\list
 {[\arabic{enumi}]}{\settowidth\labelwidth{[#1]}
 \leftmargin\labelwidth
 \advance\leftmargin\labelsep
 \usecounter{enumi}}
 \def\newblock{\hskip .11em plus .33em minus .07em}
 \sloppy
 \sfcode`\.=1000\relax}
\let\endthebibliography=\endlist
\def\lsim{\ ^<\llap{$_\sim$}\ }
\def\gsim{\ ^>\llap{$_\sim$}\ }
\def\r2{\sqrt 2}
\def\beq{\begin{equation}}
\def\eeq{\end{equation}}
\def\beqn{\begin{eqnarray}}
\def\eeqn{\end{eqnarray}}
\def\rmuu{\gamma^{\mu}}
\def\rmud{\gamma_{\mu}}
\def\PL{{1-\gamma_5\over 2}}
\def\PR{{1+\gamma_5\over 2}}
\def\sinW2{\sin^2\theta_W}
\def\AEM{\alpha_{EM}}
\def\mul{M_{\tilde{u} L}^2}
\def\mur{M_{\tilde{u} R}^2}
\def\mdl{M_{\tilde{d} L}^2}
\def\mdr{M_{\tilde{d} R}^2}
\def\mz2{M_{z}^2}
\def\c2b{\cos 2\beta}
\def\au{A_u}         
\def\ad{A_d}
\def\cob{\cot \beta}
\def\v#1{v_#1}
\def\tb{\tan\beta}
\def\epem{$e^+e^-$}
\def\KK{$K^0$-$\bar{K^0}$}
\def\wi{\omega_i}
\def\xj{\chi_j}
\def\Wmu{W_\mu}
\def\Wnu{W_\nu}
\def\m#1{{\tilde m}_#1}
\def\mH{m_H}
\def\mw#1{{\tilde m}_{\omega #1}}
\def\mx#1{{\tilde m}_{\chi^{0}_#1}}
\def\mc#1{{\tilde m}_{\chi^{+}_#1}}
\def\mwi{{\tilde m}_{\omega i}}
\def\mxi{{\tilde m}_{\chi^{0}_i}}
\def\mci{{\tilde m}_{\chi^{+}_i}}
\def\mz{M_z}
\def\sw{\sin\theta_W}
\def\cw{\cos\theta_W}
\def\cb{\cos\beta}
\def\sb{\sin\beta}
\def\rwi{r_{\omega i}}
\def\rxj{r_{\chi j}}
\def\rfp{r_f'}
\def\Kik{K_{ik}}
\def\Fq2{F_{2}(q^2)}
\def\f{\({\cal F}\)}
\def\d1{{\f(\tilde c;\tilde s;\tilde W)+ \f(\tilde c;\tilde \mu;\tilde W)}}
\def\tw{\tan\theta_W}
\def\sec2w{sec^2\theta_W}
\def\ch{{\tilde\chi^{+}_1}}
\def\c2{{\tilde\chi^{+}_2}}

\def\tt{{\tilde\theta}}

\def\tp{{\tilde\phi}}

\def\mz{M_z}
\def\sw{\sin\theta_W}
\def\cw{\cos\theta_W}
\def\cb{\cos\beta}
\def\sb{\sin\beta}
\def\rwi{r_{\omega i}}
\def\rxj{r_{\chi j}}
\def\rfp{r_f'}
\def\Kik{K_{ik}}
\def\Fq2{F_{2}(q^2)}
\def\f{\({\cal F}\)}
\def\d1{{\f(\tilde c;\tilde s;\tilde W)+ \f(\tilde c;\tilde \mu;\tilde W)}}
\def\tw{\tan\theta_W}
\def\sec2w{sec^2\theta_W}

\begin{titlepage}

\begin{center}
{\large {\bf 
 Large  Tau and Tau Neutrino Electric Dipole Moments
in Models with  Vector Like  Multiplets
}}\\
\vskip 0.5 true cm
\vspace{2cm}
\renewcommand{\thefootnote}
{\fnsymbol{footnote}}
 Tarek Ibrahim$^{a,b}$ and Pran Nath$^{b}$  
\vskip 0.5 true cm
\end{center}

\noindent
{a. Department of  Physics, Faculty of Science,
University of Alexandria,}\\
{ Alexandria, Egypt}\\ 
{b. Department of Physics, Northeastern University,
Boston, MA 02115-5000, USA} \\
\vskip 1.0 true cm

\centerline{\bf Abstract}
It is shown that the electric dipole moment of the tau lepton 
  several orders of magnitude larger than predicted by the standard model  
can be generated from mixings in models  with vector like mutiplets. 
The EDM of the  tau lepton arises from loops
involving the exchange of the W, the charginos, the neutralinos, the sleptons, 
 the mirror leptons, and the mirror  sleptons. 
The EDM of the Dirac  tau neutrino is also computed from loops
involving the exhange of the W, the charginos, the mirror leptons and 
mirror sleptons.   
A numerical analysis is presented and it is shown that the EDMs of the 
tau lepton and of the tau neutrino which lie just a couple of orders  of
magnitude below the sensitivity of the current experiment can be 
achieved. Thus the predictions of the model are testable  in improved
experiment on the EDM of the tau and of the tau neutrino.

\medskip
\end{titlepage}

\section{1. Introduction}
In the standard model the edm of the tau arises at the multiloop level and is extremely 
small\cite{Hoogeveen:1990cb}, i.e., $d_{\tau}<   10^{-34}$ecm.
On the other  hand the current experimental limit on the edm of the tau lepton\cite{Escribano:1996wp}
 is\footnote{For related papers where the upper limit on the edm of the
 tau is given to lie in the range  $10^{-16}-10^{-17}$ ecm see\cite{Inami:2002ah,Blinov:2008mu,delAguila:1990jg}.}
 \beqn
d_{\tau} < 1.1\times 10^{-17} {\rm ecm}.
\label{edmexp2}
\eeqn
Thus an experimental test of the standard model prediction is beyond the  
realm of observability in any near future experiment since the
theoretical values lies several orders of magnitude below the current experimental limits.
A similar situation also holds for the edm of the tau neutrino where 
 the current experimental limit on the edm of the tau neutrino is\cite{Escribano:1996wp}
[for related papers see\cite{GutierrezRodriguez:2004uf,Akama:2001fp}]
\beqn
d_{\tau_{\nu}} < 5.2\times 10^{-17} {\rm ecm},  
\label{edmexp1}
\eeqn
while in the standard model extended by a singlet the edm again arises only at the multiloop
level and is  many orders of magnitude below the experimental limit.
In this paper we investigate the possibility that the EDM of the tau and of the 
 tau neutrino may be much 
larger by several orders of magnitude in models where there is a small mixing of the
third generation leptons with a mirror in a vector like generation.  Such a mixing may put the tau
lepton EDM and the tau
neutrino EDM with in the realm of observation with improved experiment.
Thus vector  like combinations  are predicted in many unified  models of particle  interactions 
 \cite{Georgi:1979md,Senjanovic:1984rw}
and their implications have been explored in many recent works
\cite{Barger:2006fm,Lavoura:1992qd,Maekawa:1995ha,Liu:2009cc,Babu:2008ge,Martin:2009bg,Graham:2009gy}. Such vector like combinations
could lie in the TeV region and would be consistent with the current precision electroweak data.\\

In this work we allow for the possibility that there could be  a tiny mixing of these vector like 
combinations with the sequential generations and these mixing affect very significantly  the $\tau$
lepton moments  and also the tau
neutrino moments. 
The implications of such mixings on the magnetic moment of the tau neutrino and on the 
anomalous magnetic moment of the tau were discussed in \cite{Ibrahim:2008gg}. Here we 
include the effects of CP phases (for  a recent review of CP violation see  \cite{Ibrahim:2007fb}) and discuss the enhancement of the leptonic EDMs 
due to the mixings with mirrors in the vector like generations. 
 For the analysis here we will focus on the
leptonic vector like multiplets.
To simplify the analysis, we will assume that the mixings of the sequential generations with the 
ordinary  heavy leptons in the vector like combinations are  small and thus will ignore it. 
The inclusion of such mixings  will  affect our overall results only by a small factor $\sim 2$. 
However, we are  after much bigger effects, i.e., effects which are larger  than the SM 
results by as much as a factor of $10^{15}$.
Thus in the  following analysis we will focus on  the mixings of  the sequential generations with the 
  mirrors in the vector like combinations   and show that they have huge effects.
 The mixing of the mirrors  with the sequential leptons will introduce $V+A$ interactions for 
the ordinary leptons. Now there are very stringent constraints on such interactions  for the
first two generations and thus these mixings are effectively negligible  and we supress them
in our analysis. On the other hand for the third generation leptons, a small mixing is possible
and consistent with the current experimental constraints\cite{Singh:1987hn}.
We note  in passing that a similar situation holds for the case of the third generation
 quarks \cite{Jezabek:1994zv}.

 The masses of the vector multiplets could lie in a large mass range, i.e., from the current lower limits
  given by the  LEP experiment for  color singlet states to the region in the several TeV mass range. If the mirror leptons
  are discovered at the LHC, the analysis here would be very relevant for planning of experiments
  for the discovery of the edms of the tau and of the tau neutrino. However, it is possible that the 
  vector like multiplets have masses large enough that they might escape detection even at the LHC.
  This is specifically true for the leptonic vector multiplets since the discovery reach for them is
  typically much smaller at hadronic machines than for the color particles. However, even for this
  case the contribution of the mirrors to the edms can be huge as shown at the end of Sec.(4). 
  Specifically it shown there that with the mirror  masses in the TeV range, the edms of the tau neutrino
  and of the tau lepton can be $O(10^{14})$ larger than the Standard Model value and only
  a factor of $10^{3}$ smaller than the current sensitivity.  An improvement  in sensitivity of this
  magnitude is not necessarily outside the realm of future experiment. Further it is possible that 
  a large edm  for the tau neutrino could have astrophysical implications.

The outline of the  rest of the paper is as follows: In Sec.(2) we  give an analysis of the
 EDM of the tau lepton allowing for mixing between the vector like combination and the
third generation leptons. Here the contribution to the edm of the tau arises from the
 exchanges  of mirror neutrino, sneutrino-mirror sneutrino,
from the third generation leptons and slepton-mirror sleptons
along with W boson, chargino and neutralino exchanges.
In Sec.(3) a similar analysis is given for the EDM of the tau
neutrino with inclusion of 
the contributions arise from exchanges of the leptons from the
third generation and from the mirrors, and also from the
exchanges of the W bosons, charginos, sleptons and mirror sleptons. 

A numerical analysis of sizes of the EDM of the tau  lepton and of the tau neutrino
are given in Sec.(4). In this section we also give a display of the EMDs on the phases  
and mixings.  Conclusions are given in Sec.(5). Deductions of the mass matrices used
in Sec.(2) and Sec.(3) are given in the Appendix. 

\begin{figure}
 \vspace{0cm}
      \scalebox{1.0}
      {
       \hspace{-1cm}      
       \includegraphics[width=14cm,height=5cm]{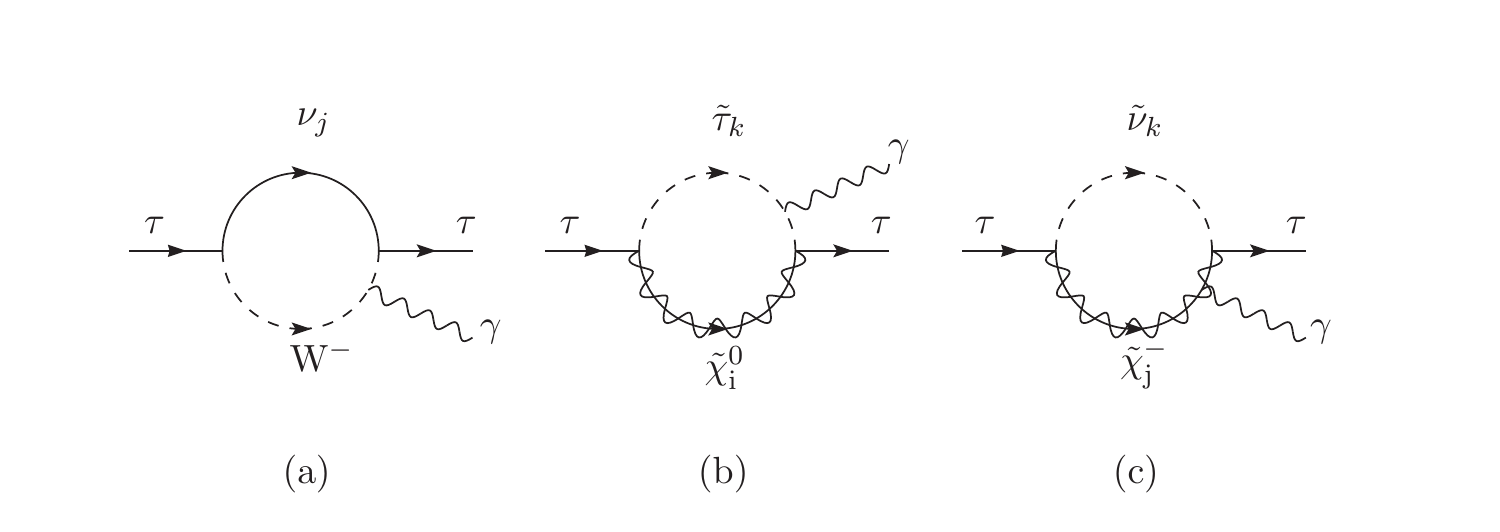}
     }     
      \vspace{0cm}
\caption{The loop contributions to the electric
dipole moment of the tau via exchange of  the $W$ boson and of tau neutrino and mirror 
neutrino denoted by $\nu_j$,
via neutralino ($\tilde \chi^0_i$) and sleptons ($\tilde \tau_k$) exchange
 and via the exchange  of 
charginos ($\tilde \chi_j^-$), sneutrinos and mirror sneutrinos denoted by ($\tilde \nu_k$).}
\label{fig:tau}
       \end{figure}

\section{2. EDM of the tau lepton}
Fig.(1a)   produces edm of the tau  ($d_{\tau}$) 
through the interaction of the W bosons with the
tau and with the neutrino and its mirror, and 
we give here  the relevant part of the Lagrangian which is 
\beqn
{\cal{L}}_{CC}=-\frac{g_2}{\sqrt 2} W^{-}_{\mu}
\sum_{j,k=1,2} \bar \tau_{k} \gamma^{\mu} 
[D^{\nu}_{L1j} D^{\tau *}_{L1k} P_L+ 
 D^{\nu}_{R2j} D^{\tau *}_{R2k} P_R] 
 \nu_{j} +H.c. 
\label{LR}
\eeqn
where  $D^{\tau, \nu}_{L,R}$ are the diagonalizing  matrices defined in the 
 Appendix.  These matrices
contain phases  and these phases
 generate  the edm of the tau  for the case of Fig.(1a).\\
 
Fig. (1b) produces a contribution to $d_{\tau}$ through neutralino exchange and the relevant interactions in
this case are
\beqn
-{\cal{L}}_{\tau - \tilde{\tau}- \chi^0}= 
\sum_{j=1-4} \sum_{k=1-4}  
\bar\tau_{1}[C_{jk}P_L+F_{jk}P_R]\tilde{\chi}^0_j \tilde \tau_k +H.c.,
\label{neutralino1}
\eeqn 
where
\beqn
C_{jk}=\sqrt{2} [\alpha_{\tau_j} D^{\tau \dagger}_{R11} \tilde D^{\tau}_{1k} 
-\gamma_{\tau_j} D^{\tau \dagger}_{R11} \tilde D^{\tau}_{3k}
+\beta_{E\tau_j} D^{\tau \dagger}_{R12} \tilde D^{\tau}_{4k}
-\delta_{E\tau_j} D^{\tau \dagger}_{R12} \tilde D^{\tau}_{2k}]\nonumber\\
F_{jk}=\sqrt{2} [\beta_{\tau_j} D^{\tau \dagger}_{L11} \tilde D^{\tau}_{1k} 
-\delta_{\tau_j} D^{\tau \dagger}_{L11} \tilde D^{\tau}_{3k}
+\alpha_{E\tau_j} D^{\tau \dagger}_{L12} \tilde D^{\tau}_{4k}
-\gamma_{E\tau_j} D^{\tau \dagger}_{L12} \tilde D^{\tau}_{2k}]
\eeqn

 Fig.(1c) produces   a  contribution to $d_{\tau}$  through chargino exchange
and the relevant interactions in this case are 
\beqn
-{\cal{L}}_{\tau - \tilde{\nu}- \chi^+}= 
\sum_{j=1-2} \sum_{k=1-4}  
\bar\tau_{1}[K_{jk}P_L+L_{jk}P_R]\tilde{\chi}^c_j \tilde \nu_k +H.c.,
\label{chargino1}
\eeqn
where
\beqn
K_{jk}=-g_2[D^{\tau\dagger}_{R 1 1} \kappa_{\tau}U^*_{j2} \tilde D^{\nu}_{1 k}
-D^{\tau\dagger}_{R 1 2} U^*_{j1} \tilde D^{\nu}_{4 k}
+D^{\tau\dagger}_{R 1 2} \kappa_{N}U^*_{j2} \tilde D^{\nu}_{2 k}],\nonumber\\
L_{jk}=-g_2[D^{\tau\dagger}_{L 1 1} \kappa_{\nu}V_{j2} \tilde D^{\nu}_{3 k}
-D^{\tau\dagger}_{L 1 1} V_{j1} \tilde D^{\nu}_{1 k}
+D^{\tau\dagger}_{L 1 2} \kappa_{E_{\tau}}V_{j2} \tilde D^{\nu}_{4 k}].
 \label{chargino2}
  \eeqn
  Here  $U$ and $V$ are the matrices  that  diagonalize the chargino mass matrix $M_C$ 
  so that 
\beq
U^* M_C V^{-1}= diag (m_{\tilde{\chi_1}}^+,m_{\tilde{\chi_2}}^+), 
\eeq
 and $\kappa_N, \kappa_{\tau}$  etc that enter Eq.(\ref{chargino2})  are defined by 
 \beqn
(\kappa_N, \kappa_{\tau})
=\frac{(m_N, m_{\tau})}{\sqrt{2} M_W \cos\beta},~
(\kappa_{E_{\tau}}, \kappa_{\nu})    =\frac{(m_{E_{\tau}}, m_{\nu})}{\sqrt{2} M_W \sin\beta}.
\eeqn
 
Using these  interactions we have 
\beqn
d^W_{\tau}=\frac{g^2_2}{32\pi^2M^2_W}\sum_{j=1,2}m_{\nu_j} Im(D^{\nu }_{L1j} D^{\tau *}_{L11}D^{\nu *}_{R2j}
D^{\tau }_{R21})I_1(\frac{m^2_{\nu_j}}{M^2_W}),\nonumber\\
d^{\chi^+}_{\tau}=-\frac{g^2_2}{16\pi^2}\sum_{j=1}^2\sum_{k=1}^4 \frac{m_{\chi^+_j}}{m^2_{\tilde \nu_k}}Im(\eta_{jk})
  A(\frac{m^2_{\chi^+_j}}{m^2_{\tilde \nu_k}}), 
~d^{\chi^0}_{\tau}=-\frac{1}{16\pi^2}\sum_{j=1}^4\sum_{k=1}^4 \frac{m_{\chi^0_j}}{m^2_{\tilde \tau_k}}Im(\zeta_{jk})
  B(\frac{m^2_{\chi^0_j}}{m^2_{\tilde \tau_k}}),  
\eeqn
where $I_1(r)$, $A(r)$ and $B(r)$ are defined as  follows
\beqn
I_1(r)=\frac{2}{(1-r)^2}[1+\frac{1}{4} r+\frac{1}{4}r^2 +\frac{3r\ln r}{2(1-r)}]\nonumber\\
A(r)=\frac{1}{2(1-r)^2}[3- r +\frac{2\ln r}{(1-r)}],
~B(r)=\frac{1}{2(1-r)^2}[1+ r +\frac{2r\ln r}{(1-r)}].
\label{AB}
\eeqn
and where
\beqn
\eta_{jk}=[-D^{\tau\dagger}_{R 1 1} \kappa_{\tau}U^*_{j2} \tilde D^{\nu}_{1 k}
+D^{\tau\dagger}_{R 1 2} U^*_{j1} \tilde D^{\nu}_{4 k}
-D^{\tau\dagger}_{R 1 2} \kappa_{N}U^*_{j2} \tilde D^{\nu}_{2 k}]\nonumber\\
\times [-D^{\tau\dagger}_{L 1 1} \kappa_{\nu}V_{j2} \tilde D^{\nu}_{3 k}
+D^{\tau\dagger}_{L 1 1} V_{j1} \tilde D^{\nu}_{1 k}
-D^{\tau\dagger}_{R 1 2} \kappa_{E_{\tau}}V_{j2} \tilde D^{\nu}_{4 k}],\nonumber\\
\zeta_{jk}=2[\alpha_{\tau j}D^{\tau\dagger}_{R 1 1} \tilde D^{\tau}_{1 k}
-\gamma_{\tau j}D^{\tau\dagger}_{R 1 1} \tilde D^{\tau}_{3 k}
+\beta_{E_{\tau j}} D^{\tau\dagger}_{R 1 2}\tilde D^{\tau}_{4 k}
-\delta_{E_{\tau j}}D^{\tau\dagger}_{R 1 2}\tilde D^{\tau}_{2 k} ]\nonumber\\
\times
[\beta^{*}_{\tau j}D^{\tau}_{L 1 1} \tilde D^{\tau *}_{1 k}
-\delta^{*}_{\tau j}D^{\tau}_{L 1 1} \tilde D^{\tau *}_{3 k}
+\alpha^{*}_{E_{\tau j}} D^{\tau }_{L 2 1}\tilde D^{\tau *}_{4 k}
-\gamma^{*}_{E_{\tau j}}D^{\tau }_{L 2 1}\tilde D^{\tau *}_{2 k} ]. 
\label{etazeta}
\eeqn
The matrix elements $\tilde D^{\nu, \tau}$ are the diagonalizing matrices of the sneutrino and slepton $4\times 4$ mass matrices (see the Appendix).
The couplings that enter $\zeta_{jk}$ in Eq.(\ref{etazeta}) are given by
\beqn\label{alphabk}
\alpha_{E_{\tau j}} =\frac{g_2 m_{E} X^*_{4j}}{2m_W\sin\beta},~~
\beta_{E_{\tau j}}=eX_{1j}^{'} +\frac{g_2}{\cos\theta_W} X_{2j}^{'}
(\frac{1}{2}-\sin^2\theta_W),\nonumber\\
\gamma_{E_{\tau j}}=e X_{1j}^{'*}-\frac{g_3\sin^2\theta_W}{\cos\theta_W}
X_{2j}^{*'},
~~ \delta_{E_{\tau j}}=-\frac{g_2 m_{E} X_{4j}}{2m_W \sin\beta}, 
\eeqn
and
\beqn
\alpha_{\tau j} =\frac{g_2 m_{\tau} X_{3j}}{2m_W\cos\beta},~~
\beta_{\tau j}=-eX_{1j}^{'*} +\frac{g_2}{\cos\theta_W} X_{2j}^{'*}
(-\frac{1}{2}+\sin^2\theta_W),\nonumber\\
\gamma_{\tau j}=-e X_{1j}'+\frac{g_2\sin^2\theta_W}{\cos\theta_W}
X_{2j}',
~~ \delta_{\tau j}=-\frac{g_2 m_{\tau} X^*_{3j}}{2m_W \cos\beta},
\eeqn
where
\beqn
X'_{1j}= (X_{1j}\cos\theta_W + X_{2j} \sin\theta_W), 
~X'_{2j}=  (-X_{1j}\sin\theta_W + X_{2j} \cos\theta_W), 
\eeqn
and where the matrix $X$ diagonlizes the neutralino mass matrix so that
\beq
X^T M_{\tilde{\chi}^0} X=diag(m_{{\chi^0}_1}, m_{{\chi^0}_2}, m_{{\chi^0}_3}, m_{{\chi^0}_4}).
\eeq

\section{3. EDM of  the tau neutrino}
The edm of the tau neutrino receives contributions from the diagrams of Fig.(2).
Using the interactions of Eq.(\ref{LR}) the contributions from the  
 loop diagrams of Fig.(2a) and Fig.(2b) are as follows:
\beqn
d^{2(a)}_{\nu}=-\frac{g^2_2}{32\pi^2M^2_W}\sum_{j=1,2}m_{\tau_j} Im(D^{\nu *}_{L11} D^{\tau}_{L1j}D^{\nu}_{R21}
D^{\tau *}_{R2j})I_1(\frac{m^2_{\tau_j}}{M^2_W})\nonumber\\
d^{2(b)}_{\nu}=-\frac{g^2_2}{32\pi^2M^2_W}\sum_{j=1,2}m_{\tau_j} Im(D^{\nu *}_{L11} D^{\tau}_{L1j}D^{\nu}_{R21}
D^{\tau *}_{R2j})I_2(\frac{m^2_{\tau_j}}{M^2_W})
\label{edm1}
\eeqn 
where $I_1(r)$ is given by Eq.(\ref{AB})  and $I_2(r)$ is given by
\beqn
I_2(r)=\frac{2}{(1-r)^2}[1-\frac{11}{4} r+\frac{1}{4}r^2 -\frac{3r^2\ln r}{2(1-r)}]
\eeqn
Similarly, the loop  
 contributions of Figs (2c) and (2d)  to $d_{\nu}$ are given by
\beqn
d^{2(c)}_{\nu}=-\frac{1}{16\pi^2}\sum_{j=1}^2\sum_{k=1}^4 \frac{m_{\chi^+_j}}{m^2_{\tilde \tau k}}
Im(S_{jk}T^*_{jk})  B(\frac{m^2_{\chi^+_j}}{m^2_{\tilde\tau k}}), \nonumber\\
d^{2(d)}_{\nu}=\frac{1}{16\pi^2}\sum_{j=1}^2\sum_{k=1}^4 \frac{m_{\chi^+_j}}{m^2_{\tilde \tau k}}Im(S_{jk}T^*_{jk})
  A(\frac{m^2_{\chi^+_j}}{m^2_{\tilde\tau k}}),
\eeqn
where $S_{jk}$ and $T_{jk}$ are given by
\beqn
S_{jk}=-g_2[D^{\nu\dagger}_{R 1 1} \kappa_{\nu}V^*_{j2} \tilde D^{\tau}_{1 k}
-D^{\nu\dagger}_{R 1 2} V^*_{j1} \tilde D^{\tau}_{4 k}
+D^{\nu\dagger}_{R 1 2} \kappa_{E_{\tau}}V^*_{j2} \tilde D^{\tau}_{2 k}],\nonumber\\
T_{jk}=-g_2[D^{\nu\dagger}_{L 1 1} \kappa_{\tau}U_{j2} \tilde D^{\tau}_{3 k}
-D^{\nu\dagger}_{L 1 1} U_{j1} \tilde D^{\tau}_{1 k}
+D^{\nu\dagger}_{L 1 2} \kappa_{N}U_{j2} \tilde D^{\tau}_{4 k}].
 \label{charginooo}
  \eeqn
and $A(r)$ and $B(r)$ are as defined in Eq.(\ref{AB}).
\begin{figure}
 \vspace{0cm}
      \scalebox{1.0}
      {
       \hspace{-1cm}      
       \includegraphics[width=18cm,height=5cm]{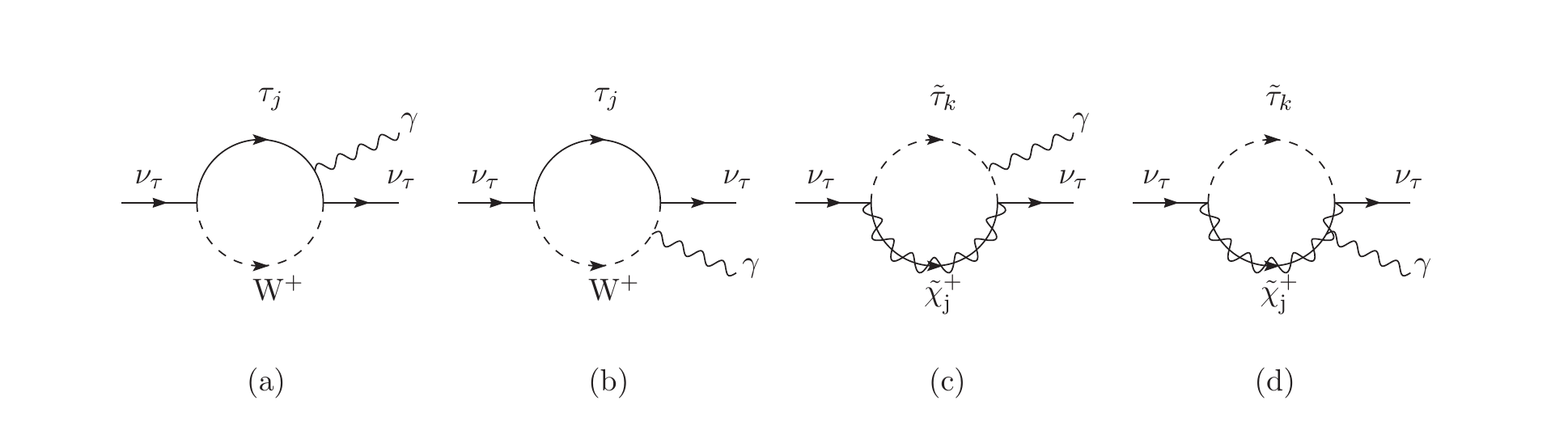}
     }     
      \vspace{0cm}
\caption{The loop contribution to the electric
dipole moment of the tau neutrino ($\nu_{\tau}$)  via exchange of the $W$ boson and of 
tau and mirror lepton denoted by $\tau_j$,  and  via exchange of the 
charginos ($\tilde\chi_j^+$), sleptons and  of mirror sleptons denoted by ($\tilde \tau_k$).}
\label{fiig:nu}
       \end{figure}

\section{4. Numerical Analysis}

The mixing matrices between leptons and mirrors are diagonalized using bi-unitary matrices (see the Appendix). 
So we parametrize the mixing between $\tau$ and $E_{\tau}$  by the angles $\theta_L$, $\theta_R$, $\chi_L$ and $\chi_R$, 
and the mixing between $\nu$ and $N$ by the angle $\phi_L$, $\phi_R$, $\xi_L$ and $\xi_R$
where 
\beqn
D^{\tau}_L=
 {\left(
\begin{array}{cc}
\cos\theta_L & -\sin\theta_L e^{-i\chi_L} \cr
             \sin\theta_L  e^{i\chi_L}& \cos\theta_L
\end{array}\right)},
~D^{\nu}_L=
 {\left(
\begin{array}{cc}
\cos\phi_L & -\sin\phi_L e^{-i\xi_L} \cr
             \sin\phi_L  e^{i\xi_L}& \cos\phi_L
\end{array}\right)},
\eeqn
and $D^{\tau}_R$ and $D^{\nu}_R$ can be gotten from $D^{\tau}_L$ and $D^{\nu}_L$
by the following substitution:  
$D^{\tau}_L\to  D^{\tau}_R, \theta_L\to \theta_R,  \chi_L\to \chi_R$,    and 
$D^{\nu}_L \to D^{\nu}_R, \phi_L\to \phi_R, \xi_L\to \xi_R$.
We note that  the phases $\chi_{L,R}$ arise from the couplings $f_3$ and $f_4$ 
while the phases $\xi_{L,R}$ arise from the couplings $f_3$ and $f_5$ through 
the relations
\beqn
\chi_R=arg (m_{\tau}f_3+m_Ef^*_4),
~\chi_L=arg (m_{\tau}f^*_4+m_E f_3),\nonumber\\
~\xi_R=arg (-m_{\nu}f_3+m_Nf^*_5),
~\xi_L=arg (m_{\nu}f^*_5-m_N f_3).
\eeqn
However, these four parameters are not independent
since the input of three phases of $f_3$, $f_4$ and $f_5$ would produce these
four parameters.
For the case of lepton and neutrino masses arising from hermitian matrices, i.e., when
 $f_4=f^*_3$ and $f_5=-f^*_3$ we have
$\theta_L=\theta_R$, $\phi_L=\phi_R$, $\chi_L=\chi_R=\chi$ and $\xi_L=\xi_R=\xi$.  Further,   here we have
the relation $\xi=\chi+\pi$ and thus the W-exchange terms of the edms for tau neutrinos and tau leptons
vanish.  However, more generally  the lepton and the neutrino mass matrices are not hermitian and  they generate 
non-vanishing  contributions to the EDMs.
Thus the input parameters for this sector of the parameter space are
$m_{\tau 1},m_E, f_3, f_4, m_{\nu 1}, m_N, f_5$
with $f_3$, $f_4$ and $f_5$ being  complex masses with CP violating phases $\chi_3$, $\chi_4$ and 
$\chi_5$ respectively.
 For the slepton mass$^2$ matrices we need the extra input parmaeters of the susy breaking sector,
$
\tilde{M}_L, \tilde{M}_E,\tilde{M}_{\tau},\tilde{M}_{\chi},\tilde{M}_{\nu},\tilde{M}_N,
A_{\tau}, A_E, A_{\nu},A_{N}, \mu, \tan\beta.
$
The chargino and neutralino sectors need the extra two parameters
$\tilde{m}_1, \tilde{m}_2$.
We will assume that the only parameters that have phases in the above set are 
$A_E$, $A_N$, $A_{\tau}$ and $A_{\nu}$.
These phases are $\alpha_E$, $\alpha_N$, $\alpha_{\tau}$ and $\alpha_{\nu}$ respectively.
To simplify the analysis we set the phases $\alpha_{\nu}=\alpha_{\tau}=0$. With this in mind
the only contributions to the edm of the tau lepton and tau neutrino arise from mixing terms between the scalar
matter - scalar mirrors, fermion matter - fermion mirror and finally between mirrors among themselves in the scalar
sector.
Thus in the absense of mirror part of the lagrangian, the edms of taus and neutrinos vanish. We can thus isolate
the  role of the CP violating phases in this sector and see the size of its contribution.
The $4\times 4$ mass$^2$ matrices of sleptons and sneutrinos are diagonlized numerically.
Thus the CP violating phases that would play a role in this analysis are
\beq
\chi_3, \chi_4, \chi_5, \alpha_E, \alpha_N.
\eeq
To reduce the number of input parameters we assume
$\tilde{M}_a =m_0, a=L, E, \tau, \chi, \nu, N$ and $|A_i|=|A_0|$, $i=E, N, \tau, \nu$.

In  Fig.({\ref{fig3}), we  give a numerical analysis of the edm of the tau lepton and discuss
its  variation with the parameter $\chi_3$  (left), with $|f_3|$) (middle) and with $\alpha_N$ (right). 
Regarding $\chi_3$ it enters $D^{\nu}$, $D^{\tau}$, $\tilde D^{\nu}$ and $\tilde D^{\tau}$ and as a consequences all diagrams
in Fig.(1) that contribute to the edm of the tau  are affected. The phase $\alpha_N$, however,
enters only in the chargino exchange contribution since it enters $\tilde D^{\nu}$ and thus only the
part of the  tau  edm arising from the chargino exchanged is affected by variations of
$\alpha_N$.  We note that the various diagrams Figs.(1a)- Fig.(1c) that contribute to the tau
 edm can add constructively or destructively in the latter case generating large
cancellations reminiscent of the cancellation mechanism for the edm of the electron and for 
the neutron\cite{Ibrahim:2007fb}.  
Of course the desirable larger contributions  for the tau  edm occur away
from the cancellation regions. The analysis of Fig.(\ref{fig3}) shows that a tau edm 
as large $10^{-18}- 10^{-19}$ecm  can be gotten which is only about 2 orders of magnitude 
below the current experimental limits of Eq.(\ref{edmexp2}). 
A similar analysis for the tau neutrino edm is given in Fig.(\ref{fig4}).  Here again one  finds  that the
tau neutrino edm as large as $10^{-18}-10^{-19}$ ecm can be gotten and  again it lies only a couple of
orders of magnitude below the current experimental limit of Eq.(\ref{edmexp1}).

  As discussed in Sec.(1) the scale of the vector like multiplets is unknown. 
 They could lie in the sub TeV region  but on the
 other hand they could also be several TeV size and escape direct detection even at the LHC.
 This is especially true for leptonic  vector like multiplets since the discovery reach for color singlet leptonic
 states at  hadronic machines is typically much smaller than for the color particles. In this context it is
 then interesting to investigate the contributions to the tau lepton edm and to the tau neutrino edm
 from vector like leptonic multiplets in the TeV range.  A comparison of these edms when the 
 leptonic vector like multiplet lies in the sub TeV region vs in the TeV region is given in 
 Table 1 below. It turns out that the dependence of the  edms on the masses of the mirror leptons
 is a rather complicated one.  Thus there are supersymmetric and non -supersymmetric contributions
 which have different dependence on the mirrors  and mirror slepton masses.  Thus in certain
 regions of the parameter space as the mirror leptons masses  
grow, the chargino and neutralino contributions are suppressed much faster
than the W-exchange terms. In the susy contributions, both the couplings
and form factors that contain the mirror lepton masses explicitly decrease
as the mirror spectrum increases. In the W-exchange term, there is a
competition between the couplings and form factors. The first term decreases while 
the latter increases and,  in indeed a suppression of the edms occurs but here
it is much slower than in the case of susy case.  These phenomena are  illustrated in the
analysis of Table 1 which shows that the suppression of the W exchange terms in both tau 
and neutrino edms is much slower rate than the other components in this specific 
part of the parameter space. 
\begin{center} \begin{tabular}{|c|c|c|c|c|c|c|}
\multicolumn{7}{c} {Table~1:  } \\
\hline
$m_E (TeV)$ & $m_N (TeV)$  &$d^W_{\tau}e.cm$ & $d^{\chi^+}_{\tau}e.cm$ &
$d^{\chi^0}_{\tau}e.cm$ & $d^W_{\nu}e.cm$ & $d^{\chi^+}_{\nu}e.cm$ \\
\hline
\hline
$0.1$     &  $0.2$  &  $6.5\times 10^{-18}$     &   $-3.4\times 10^{-18} $
&    $5.0\times 10^{-19}$ & $3.7\times 10^{-18}$
&$-2.4\times 10^{-18}$\\
 \hline
$2.0$     &  $1.0$  &  $4.0\times 10^{-20}$     &   $-7.2\times 10^{-22}$
&    $3.0\times 10^{-23} $& $5.1\times 10^{-20}$
&$-7.1\times 10^{-22}$ \\
 \hline
\end{tabular}\\~\\
\label{tab:1}
\noindent
\end{center}
Table caption:  A sample illustration of the
contributions to the electric dipole moments of  $\nu_{\tau}$
 and of $\tau$.  The in puts are:  $\tan\beta=$5, $|f_3|=$90, $|f_4|=$120,
$|f_5|=$75, $m_0=$150, $|A_0|=$100, $\tilde{m}_1=75$, $\tilde{m}_2=150$,
$\mu=130$,
$\chi_3=-$1.0, $\chi_4=$0.6, $\chi_5=-$0.8, $\alpha_E=$0.3 and
$\alpha_N=$0.6.
All masses are in units of GeV and all angles are in radian.

The analysis given above  shows that even if the vector like particles lie in the 
TeV range they could  contribute a significant amount, i.e., $O(10^{-20}) ecm$ 
which is $O(10^{14})$ larger than what the Standard Model predicts and 
only three  order  of magnitude  smaller than the current limits. The above results
do not appear outside the realm of detection in future experiment with improved
sensitivity. Further, the results above could have possible astrophysical 
implications.


\section{5. Conclusion}
In this paper we have  considered extensions of the MSSM with vector like multiplets.
We have specifically focused on the leptonic sector and considered mixings between the 
sequential generation leptons and  the mirrors in the vector like multiplets. For the first
two generations of leptons the $V-A$ structure of the weak interactions are very well 
established. However, this is not the case for the third generation leptons.  Thus for the
third generation leptons we consider small mixings of the tau lepton and of the tau
neutrino with the mirrors in the vector like generation. An analysis of the electric dipole
moment of the tau lepton is carried out in this framework. Further, we also compute 
the EDM of the tau neutrino.  It is found that the predictions of the EDMs in the model can be  as large
as just a couple of  orders of magnitude below the current experimental predictions.  Thus an
improvement  in experiment by this  order of magnitude  will begin to test the 
predictions of the model.  These  results are  very encouraging for the possible 
observation of the EDM of the tau lepton and of the EDM of the tau neutrino in improved
experiment.

{\em Acknowledgments}:  
This research is  supported in part by  NSF grant PHY-0757959 and by PHY-0704067.

\section{ Appendix: Mass Matrices of leptons and sleptons and their mirrors}
In this appendix we write down the mass matrices for the leptons, neutrinos and sleptons 
and their mirrors  that
enter in the computations of the edms of the tau neutrino and tau lepton discussed in the text
of the paper.
In deducing 
these matrices we need the transformation properties of the leptons and  their mirrors.
Thus under $SU(3)_C\times SU(2)_L \times U(1)_Y$ the leptons transform as  follows
\beqn
\psi_L\equiv \left(
\begin{array}{c}
 \nu_L\\
 \tau_L
\end{array}\right) \sim(1,2,- \frac{1}{2}), \tau^c_L\sim (1,1,1), \nu^c_L\sim (1,1,0),
\eeqn
where the last entry on the right hand side of each $\sim$ is the value of the hypercharge
 $Y$ defined so that $Q=T_3+ Y$ and we have included in our analysis the singlet field 
 $\nu^c$.
  These leptons have $V-A$ interactions. Let us now consider mirror
leptons in the vector like mutiplets  which have $V+A$ interactions
(For previous works on mirrors see \cite{Maalampi:1988va}). Their quantum numbers are
as follows
\beqn
\chi^c\equiv \left(
\begin{array}{c}
 E_{\tau L}^c\\ 
 N_L^c
\end{array}\right)
\sim(1,2,\frac{1}{2}), E_{\tau L}\sim (1,1,-1), N_L\sim (1,1,0).
\eeqn
We assume that the mirrors of the vector like generation escape acquiring mass at the GUT scale and remain
light down to the electroweak scale where the superpotential of the model
for the lepton part  may be written  in the form
\beqn
W= \epsilon_{ij}  [f_{1} \hat H_1^{i} \hat \psi_L ^{j}\hat \tau^c_L
 +f_{1}' \hat H_2^{j} \hat \psi_L ^{i} \hat \nu^c_L
+f_{2} \hat H_1^{i} \hat \chi^c{^{j}}\hat N_{L}
 +f_{2}' \hat H_2^{j} \hat \chi^c{^{i}} \hat E_{\tau L}]\nonumber\\
+ f_{3} \epsilon_{ij}  \hat \chi^c{^{i}}\hat\psi_L^{j}
 + f_{4} \hat \tau^c_L \hat  E_{\tau L}  +  f_{5} \hat \nu^c_L \hat N_{L}.
\label{superpotential}
\eeqn
Mixings of the above type can arise 
 via non-renormalizable interactions\cite{Senjanovic:1984rw}.
Consider, for example,  a term  such as 
$1/M_{Pl} \nu^c_LN_L \Phi_1\Phi_2$. If $\Phi_1$ and $\Phi_2$ develop VEVs of size $10^{9-10}$,
a mixing term of the right size can be generated.

To get the mass matrices of the leptons and of the mirror leptons we 
replace the superfields in the superpotential by their component scalar
fields. The relevant parts in the superpotential that produce the lepton and
mirror lepton mass matrices are
\beqn
W=f_1 H_1^1 \tilde{\tau}_L \tilde{\tau}_R^* +f_1' H_2^2 \tilde{\nu}_{L} \tilde{\nu}^*_{R}+
f_2 H_1^1 \tilde{N}_R^* \tilde{N}_L+f_2' H_2^2 \tilde{E}^*_{\tau R} \tilde{E}_{\tau L}\nonumber\\
+f_3 \tilde{E}^*_{\tau R} \tilde{\tau}_L -f_3 \tilde{N}_R^* \tilde{\nu}_{L}+ f_4 \tilde{\tau}_R^* \tilde{E}_{\tau L}
+f_5 \tilde{\nu}^*_{R} \tilde{N}_L.
\eeqn
The mass terms for the lepton and their mirrors arise from the part of the lagrangian
\beq
{\cal{L}}=-\frac{1}{2}\frac{\partial ^2 W}{\partial{A_i}\partial{A_j}}\psi_ i \psi_ j+H.c.
\eeq
where $\psi$ and $A$ stand for generic two-component fermion and scalar fields.
After spontaneous breaking of the electroweak symmetry, ($<H_1^1>=v_1/\sqrt{2} $ and $<H_2^2>=v_2/\sqrt{2}$),
we have the following set of mass terms written in 4-spinors for the fermionic sector 
\beqn
-{\cal L}_m = \br\bar \tau_R ~ \bar E_{\tau R} \er
 \br
  f_1 v_1/\sqrt{2} ~~ f_4\\
 f_3 ~~ f_2' v_2/\sqrt{2}\er
 \br \tau_L\\
 E_{\tau L}\er
  + \br\bar \nu_R ~~ \bar N_R\er
 \br f'_1 v_2/\sqrt{2}~~ f_5\\
 -f_3 ~~ f_2 v_1/\sqrt{2}\er \br \nu_L\\
 N_L\er  + H.c.\nonumber
\eeqn 
Here 
the mass matrices are not  Hermitian and one needs
to use bi-unitary transformations to diagonalize them. Thus we write the linear transformations
\beqn
 \br\tau_R\\ 
 E_{\tau R}\er=D^{\tau}_R \br\tau_{1_R}\\
 E_{\tau 2_R} \er,
~\br \tau_L\\
 E_{\tau L}\er=D^{\tau}_L \br \tau_{1_L}\\
 E_{\tau 2_L}\er,
\eeqn
such that
\beq
D^{\tau \dagger}_R \br f_1 v_1/\sqrt{2} ~~ f_4\\
 f_3 ~~ f_2' v_2/\sqrt{2}\er D^{\tau}_L=diag(m_{\tau_1},m_{\tau_2}),
\label{put1}
\eeq
and the same holds for the neutrino mass matrix so that 
\beq
D^{\nu \dagger}_R \br f'_1 v_2/\sqrt{2} ~~ f_5\\
 -f_3 ~~ f_2v_1/\sqrt{2}\er D^{\nu}_L=diag(m_{\nu_1},m_{\nu_2}).
\label{put2}
\eeq
Here $\tau_1, \tau_2$ are the mass eigenstates and we identify the tau lepton 
with the eigenstate 1, i.e.,  $\tau=\tau_1$, and identify $\tau_2$ with a heavy 
mirror eigenstate  with a mass in the hundreds  of GeV. Similarly 
$\nu_1, \nu_2$ are the mass eigenstates for the neutrinos, 
where we identify $\nu_1$ with the light neutrino state and $\nu_2$ with the 
heavier mass eigen state.
By multiplying Eq.(\ref{put1}) by $D^{\tau \dagger}_L$ from the right and by
$D^{\tau}_R$ from the left and by multiplying Eq.(\ref{put2}) by $D^{\nu \dagger}_L$
from the right and by $D^{\nu}_R$ from the left, one can equate the values of the parameter
$f_3$ in both equations and we can get the following relation
between the diagonalizing matrices $D^{\tau}$ and $D^{\nu}$
\beq
m_{\tau 1} D^{\tau}_{R 21} D^{\tau *}_{L 11} +m_{\tau 2} D^{\tau}_{R 22} D^{\tau *}_{L 12}=
-[m_{\nu 1} D^{\nu}_{R 21} D^{\nu *}_{L 11} +m_{\nu 2} D^{\nu}_{R 22} D^{\nu *}_{L 12}].
\label{condition}
\eeq

Next we  consider  the mixings of the charged sleptons and the charged mirror sleptons. 
We write the superpotential in terms of the scalar fields of interest as follows
\beqn
W= -\mu \epsilon_{ij} H_1^i H_2^j+\epsilon_{ij}  [f_{1}  H_1^{i} \tilde \psi_L ^{j}\tilde \tau^c_L
 +f_{1}'  H_2^{j} \tilde \psi_L ^{i} \tilde \nu^c_L
+f_{2}  H_1^{i} \tilde \chi^c{^{j}}\tilde N_{L}
 +f_{2}'  H_2^{j} \tilde \chi^c{^{i}} \tilde E_{\tau L}]\nonumber\\
+ f_{3} \epsilon_{ij}  \tilde \chi^c{^{i}}\tilde \psi_L^{j}
 + f_{4} \tilde \tau^c_L \tilde  E_{\tau L}  +  f_{5} \tilde \nu^c_L \tilde N_{L}.
\label{superpotentials}
\eeqn
The mass$^2$ matrix of the slepton - mirror slepton comes from three sources, the F term, the
D term of the potential and the soft susy breaking terms.
Using the above superpotential and after the breaking of  the electroweak symmetry we get for the
mass part of the lagrangian $ {\cal L}_F$ and $ {\cal L}_D$ the following set of terms 
\beqn
-{\cal L}_F=(m^2_E +|f_3|^2)\tilde E_R \tilde E^*_R +(m^2_N +|f_3|^2)\tilde N_R \tilde N^*_R
+(m^2_E +|f_4|^2)\tilde E_L \tilde E^*_L\nonumber\\
 +(m^2_N +|f_5|^2)\tilde N_L \tilde N^*_L
+(m^2_{\tau} +|f_4|^2)\tilde \tau_R \tilde \tau^*_R +(m^2_{\nu} +|f_5|^2)\tilde \nu_R \tilde \nu^*_R
+(m^2_{\tau} +|f_3|^2)\tilde \tau_L \tilde \tau^*_L\nonumber\\ +(m^2_{\nu} 
+|f_3|^2)\tilde \nu_L \tilde \nu^*_L
+\{-m_{\tau} \mu^* \tan\beta \tilde \tau_L \tilde \tau^*_R -m_{N} \mu^* \tan\beta \tilde N_L \tilde N^*_R 
-m_{\nu} \mu^* \cot\beta \tilde \nu_L \tilde \nu^*_R\nonumber\\
 -m_{E} \mu^* \cot\beta \tilde E_L \tilde E^*_R +(m_E f^*_3 +m_{\tau} f_4) \tilde E_L \tilde \tau^*_L 
+(m_E f_4 +m_{\tau} f^*_3) \tilde E_R \tilde \tau^*_R\nonumber\\
+(m_{\nu} f_5 -m_{N} f^*_3) \tilde N_L \tilde \nu^*_L
+(m_{N} f_5 -m_{\nu} f^*_3) \tilde N_R \tilde \nu^*_R
+h.c. \},
\eeqn
and
\beqn
-{\cal L}_D=\frac{1}{2} m^2_Z \cos^2\theta_W \cos 2\beta \{\tilde \nu_L \tilde \nu^*_L -\tilde \tau_L \tilde \tau^*_L 
+\tilde E_R \tilde E^*_R -\tilde N_R \tilde N^*_R\}\nonumber\\
+\frac{1}{2} m^2_Z \sin^2\theta_W \cos 2\beta \{\tilde \nu_L \tilde \nu^*_L +\tilde \tau_L \tilde \tau^*_L 
-\tilde E_R \tilde E^*_R -\tilde N_R \tilde N^*_R +2 \tilde E_L \tilde E^*_L -2 \tilde \tau_R \tilde \tau^*_R\}.
\eeqn
Next we add the general set of soft supersymmetry breaking terms to the scalar potential so that 
\beqn
V_{soft}=\tilde M^2_L \tilde \psi^{i*}_L \tilde \psi^i_L +\tilde M^2_{\chi} \tilde \chi^{ci*} \tilde \chi^{ci}
+\tilde M^2_{\nu} \tilde \nu^{c*}_L \tilde \nu^c_L 
+\tilde M^2_{\tau} \tilde \tau^{c*}_L \tilde \tau^c_L +\tilde M^2_E \tilde E^*_L \tilde E_L
 + \tilde M^2_N \tilde N^*_L \tilde N_L \nonumber\\
+\epsilon_{ij} \{f_1 A_{\tau} H^i_1 \tilde \psi^j_L \tilde \tau^c_L 
-f'_1 A_{\nu} H^i_2 \tilde \psi ^j_L \tilde \nu^c_L
+f_2 A_N H^i_1 \tilde \chi^{cj} \tilde N_L
-f'_2 A_E H^i_2 \tilde \chi^{cj} \tilde E_L +h.c.\}
\eeqn
From ${\cal L}_{F,D}$ and
by giving the neutral Higgs their vacuum expectation values in  $V_{soft}$ we can produce the
the mass$^2$  matrix $M^2_{\tilde \tau}$  in the basis $(\tilde  \tau_L, \tilde E_L, \tilde \tau_R, 
\tilde E_R)$. We  label the matrix  elements of these as $(M^2_{\tilde \tau})_{ij}= M^2_{ij}$ where
\beqn
M^2_{11}=\tilde M^2_L +m^2_{\tau} +|f_3|^2 -m^2_Z cos 2 \beta (\frac{1}{2}-\sin^2\theta_W), \nonumber\\
M^2_{22}=\tilde M^2_E +m^2_{E} +|f_4|^2 +m^2_Z cos 2 \beta \sin^2\theta_W, \nonumber\\
M^2_{33}=\tilde M^2_{\tau} +m^2_{\tau} +|f_4|^2 -m^2_Z cos 2 \beta \sin^2\theta_W, \nonumber\\
M^2_{44}=\tilde M^2_{\chi} +m^2_{E} +|f_3|^2 +m^2_Z cos 2 \beta (\frac{1}{2}-\sin^2\theta_W), \nonumber\\
M^2_{12}=M^{2*}_{21}=m_E f^*_3 +m_{\tau} f_4,
M^2_{13}=M^{2*}_{31}=m_{\tau}(A^*_{\tau} -\mu \tan\beta),\nonumber\\
M^2_{14}=M^{2*}_{41}=0,
M^2_{23}=M^{2*}_{32}=0,\nonumber\\
M^2_{24}=M^{2*}_{42}=m_E(A^*_E -\mu \cot \beta),
M^2_{34}=M^{2*}_{43}=m_E f_4 +m_{\tau} f^*_3.
\eeqn

Here the terms $M^2_{11}, M^2_{13}, M^2_{31}, M^2_{33}$ arise from soft  
breaking in the  sector $\tilde \tau_L, \tilde \tau_R$. Similarly the terms 
$M^2_{22}, M^2_{24},$  $M^2_{42}, M^2_{44}$ arise from soft  
breaking in the  sector $\tilde E_L, \tilde E_R$. The terms $M^2_{12}, M^2_{21},$
$M^2_{23}, M^2_{32}$, $M^2_{14}, M^2_{41}$, $M^2_{34}, M^2_{43},$  arise  from mixing between the staus  and 
the mirrors.  We assume that all the masses are of the electroweak scale 
so all the terms enter in the mass$^2$ matrix.  We diagonalize this hermitian mass$^2$ matrix  by the
 unitary transformation 
$
 \tilde D^{\tau \dagger} M^2_{\tilde \tau} \tilde D^{\tau} = diag (m^2_{\tilde \tau_1},  
m^2_{\tilde \tau_2}, m^2_{\tilde \tau_3},  m^2_{\tilde \tau_4})$.
There is a  similar mass$^2$  matrix in the sneutrino sector.
In the basis $(\tilde  \nu_L, \tilde N_L, \tilde \nu_R, \tilde N_R)$ 
we can write the sneutrino mass$^2$ matrix in the form 
$(M^2_{\nu})_{ij}=m^2_{ij}$ where
\beqn
m^2_{11}=\tilde M^2_L +m^2_{\nu} +|f_3|^2 +\frac{1}{2}m^2_Z cos 2 \beta,  \nonumber\\
m^2_{22}=\tilde M^2_N +m^2_{N} +|f_5|^2,
~m^2_{33}=\tilde M^2_{\nu} +m^2_{\nu} +|f_5|^2,  \nonumber\\
m^2_{44}=\tilde M^2_{\chi} +m^2_{N} +|f_3|^2 -\frac{1}{2}m^2_Z cos 2 \beta, 
~m^2_{12}=m^{2*}_{21}=m_{\nu} f_5 -m_{N} f^*_3,\nonumber\\
m^2_{13}=m^{2*}_{31}=m_{\nu}(A^*_{\nu} -\mu \cot\beta),
~m^2_{14}=m^{2*}_{41}=0,
~m^2_{23}=m^{2*}_{32}=0,\nonumber\\
m^2_{24}=m^{2*}_{42}=m_N(A^*_N -\mu \tan \beta),
~m^2_{34}=m^{2*}_{43}=m_N f_5 -m_{\nu} f^*_3.
\eeqn

As in the charged  slepton sector 
here also the terms $m^2_{11}, m^2_{13}, m^2_{31}, m^2_{33}$ arise from soft  
breaking in the  sector $\tilde \nu_L, \tilde \nu_R$. Similarly the terms 
$m^2_{22}, m^2_{24},$  $m^2_{42}, m^2_{44}$ arise from soft  
breaking in the  sector $\tilde N_L, \tilde N_R$. The terms $m^2_{12}, m^2_{21},$
$m^2_{23}, m^2_{32}$, $m^2_{14}, m^2_{41}$, $m^2_{34}, m^2_{43},$  arise  
from mixing between the physical sector and 
the mirror sector.  Again as in the charged lepton sector 
we assume that all the masses are of the electroweak size
so all the terms enter in the mass$^2$ matrix.  This mass$^2$  matrix can be diagonalized  by the
 unitary transformation 
$
 \tilde D^{\nu\dagger} M^2_{\tilde \nu} \tilde D^{\nu} = diag (m^2_{\tilde \nu_1},  
m^2_{\tilde \nu_2}, m^2_{\tilde \nu_3},  m^2_{\tilde \nu_4})
$.
The physical tau and neutrino states are $\tau\equiv \tau_1, \nu\equiv \nu_1$,
and the states $\tau_2, \nu_2$ are heavy states with mostly mirror particle content. 
The states $\tilde \tau_i, \tilde \nu_i; ~i=1-4$ are the slepton and sneutrino states. 
For the case of  
no mixing these limits are as  follows: 
$
\tilde \tau_1\to \tilde \tau_L, ~\tilde \tau_2\to \tilde E_L, ~\tilde \tau_3\to \tilde \tau_R, ~
\tilde \tau_4\to \tilde E_R,
~\tilde \nu_1\to \tilde \nu_L, ~\tilde \nu_2\to \tilde N_L, ~\tilde \nu_3\to \tilde \nu_R, ~
\tilde \nu_4\to \tilde N_R
$.
The couplings $f_3$, $f_4$ and $f_5$ can be complex and thus the matrices $D^{\tau}_{L,R}$ and
$D^{\nu}_{L,R}$ will have complex elements that would produce electric dipole moments through
their arguments discussed in the text  of the paper. 
Also the trilinear couplings $A_{\nu, \tau, E, N}$ could be
complex and produce electric dipole moment through the arguments of $\tilde D^{\nu}$ and $\tilde D^{\tau}$.
We will assume for simplicity that this is the only part in the theory that has
CP violating phases (For  a recent review of CP violation see  \cite{Ibrahim:2007fb}).
Thus the $\mu$ parameter is considered real along
with the other trilinear couplings in the theory. 
In this way we can automatically satisfy the constraints on the edms of  the electron,  the neutron
and of  Hg and of Thallium.

\clearpage
\begin{figure}
      \scalebox{.6}
      {
       \hspace{-2cm}      
       \includegraphics[width=10cm,height=12cm]{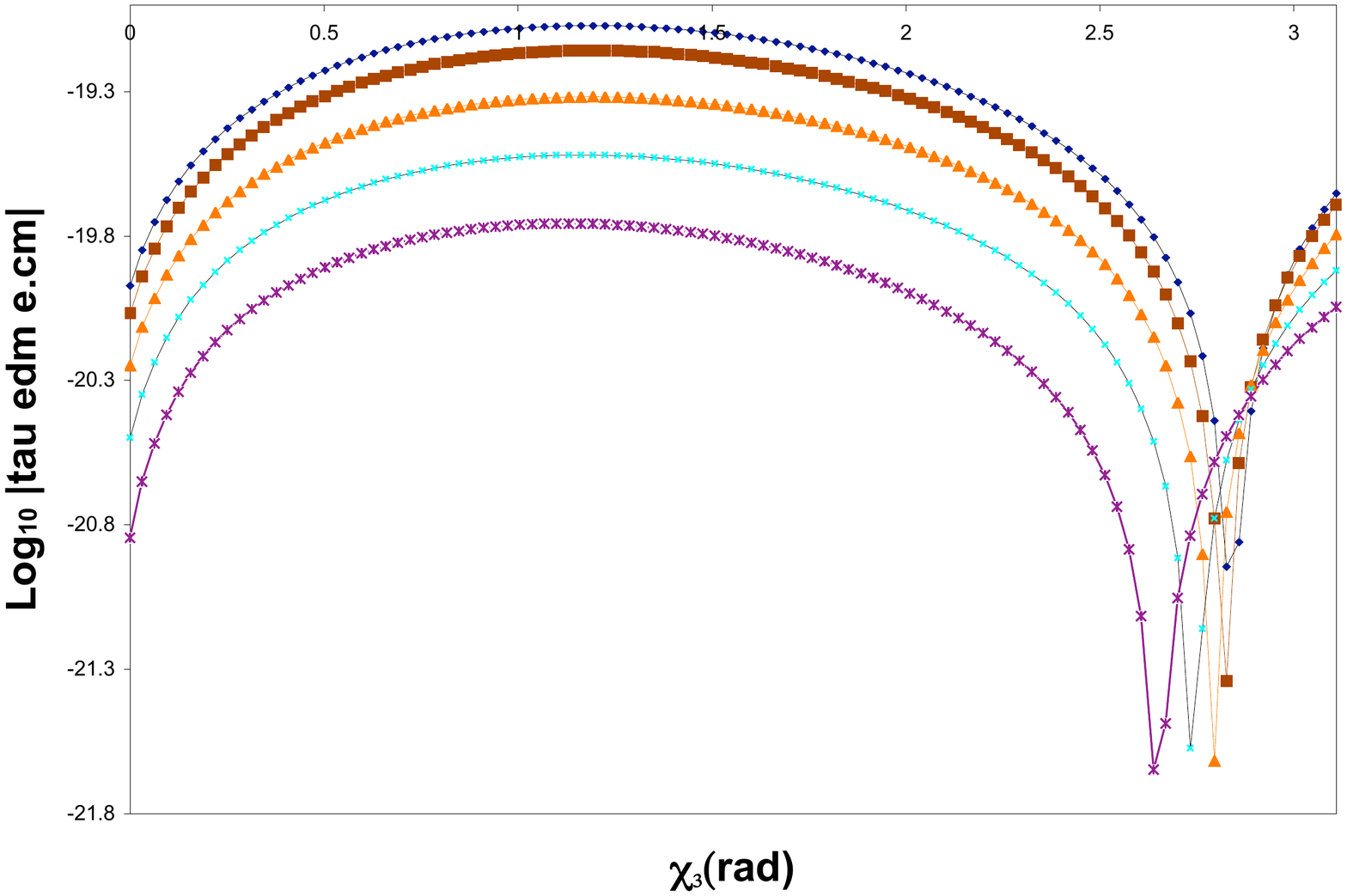}
 \includegraphics[width=10cm,height=12cm]{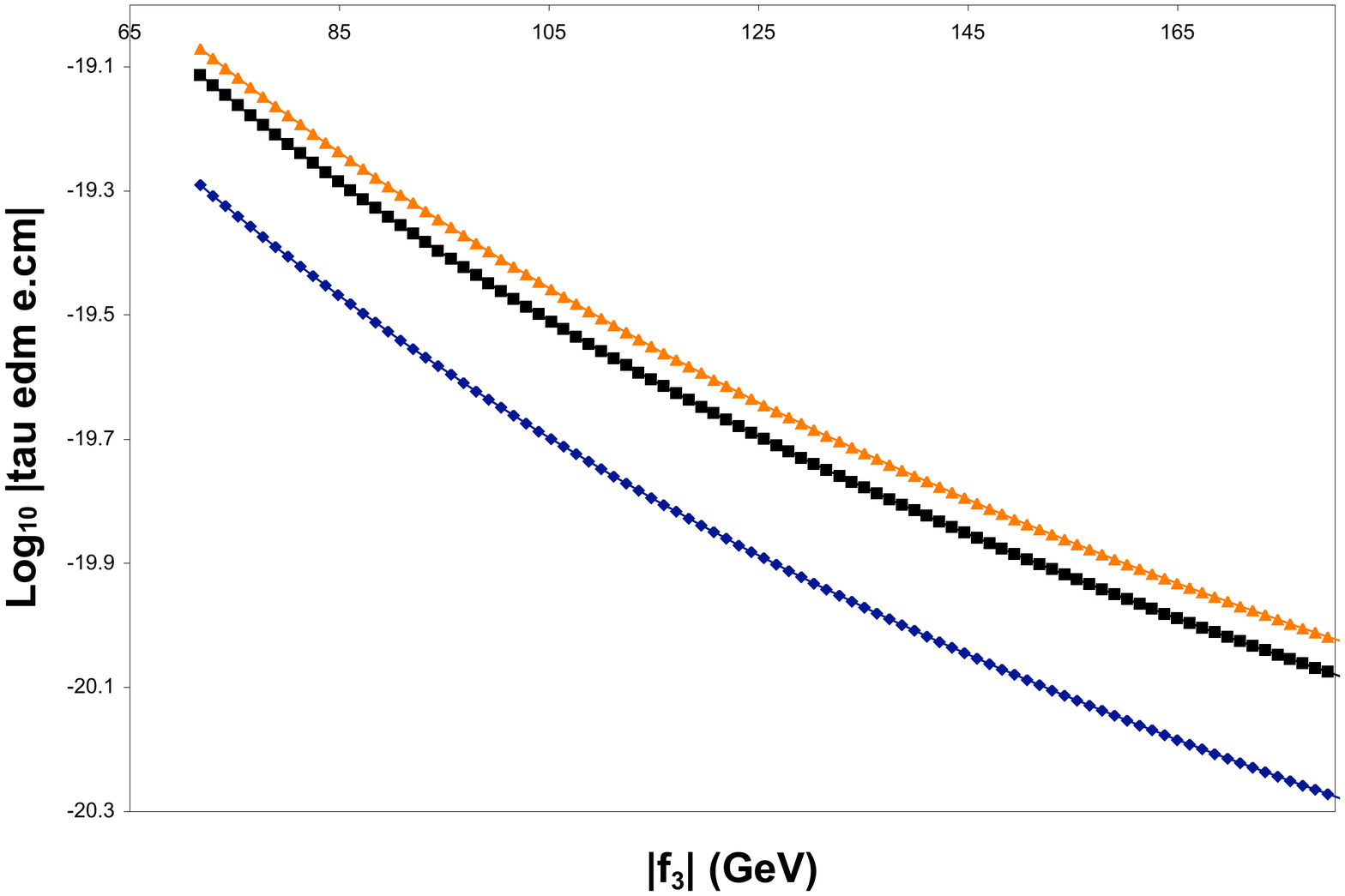}
  \includegraphics[width=10cm,height=12cm]{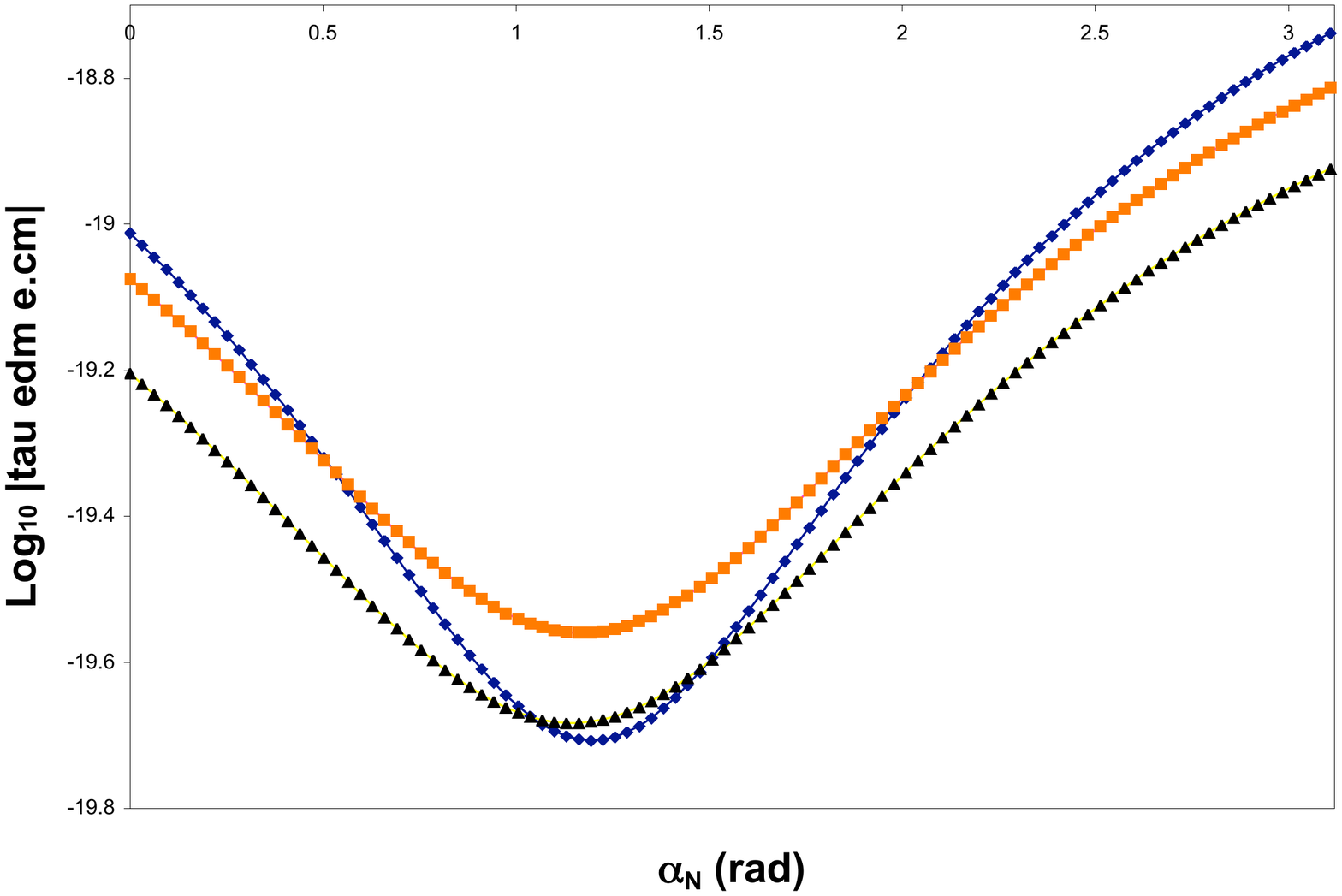}     }     
\caption{ 
\scriptsize
Left:  An exhibition of the dependence of $d_{\tau}$ on $\chi_3$ when $\tan\beta=10$, 
$m_N=100$, $|f_3|=$50, $|f_4|=$70,
$|f_5|=$90, $m_0=$100, $|A_0|=$150, $\tilde{m}_1=50$, $\tilde{m}_2=100$, $\mu=150$, $\chi_4=$0.4, $\chi_5=$0.6, $\alpha_E=$0.5,  $\alpha_N=$0.8,  
 and  $m_E=$300, 250, 200, 150, 100 (in ascending order at $\chi_3=0$).
Middle: 
An exhibition of the dependence of $d_{\tau}$ on $|f_3|$ when $\tan\beta=10$,
$m_N=$120, $m_E=$100, $|f_4|=$80,
$|f_5|=$60, $m_0=$150, $|A_0|=$100, $\tilde{m}_1=50$, $\tilde{m}_2=100$, $\mu=150$ GeV 
$\chi_4=$0.3, $\chi_5=$0.7, $\alpha_E=$0.4, $\alpha_N=$1.0, $\alpha_{\tau}=0$, $\alpha_{\nu}=0$, and  $\chi_3=$0.4, 0.8, 1.2 (in ascending order).
Right: An exhibition of the dependence of $d_{\tau}$ on $\alpha_N$ when $\tan\beta=20$
$m_N=100$, $|f_3|=$70, $|f_4|=$50,
$|f_5|=$80, $m_0=$120, $|A_0|=$130, $\tilde{m}_1=50$, $\tilde{m}_2=100$, $\mu=150$, $\chi_3=$0.5, $\chi_4=$0.6, $\chi_5=$0.7 and $\alpha_E=$0.6,  and $m_E=$180, 130, 80 (in ascending order at $\alpha_N=0$). 
Masses in GeV and  angles in rad here and in figures below.
}
\label{fig3}
       \end{figure}

\begin{figure}
      \scalebox{.6}
      {
       \hspace{-2cm}      
       \includegraphics[width=10cm,height=12cm]{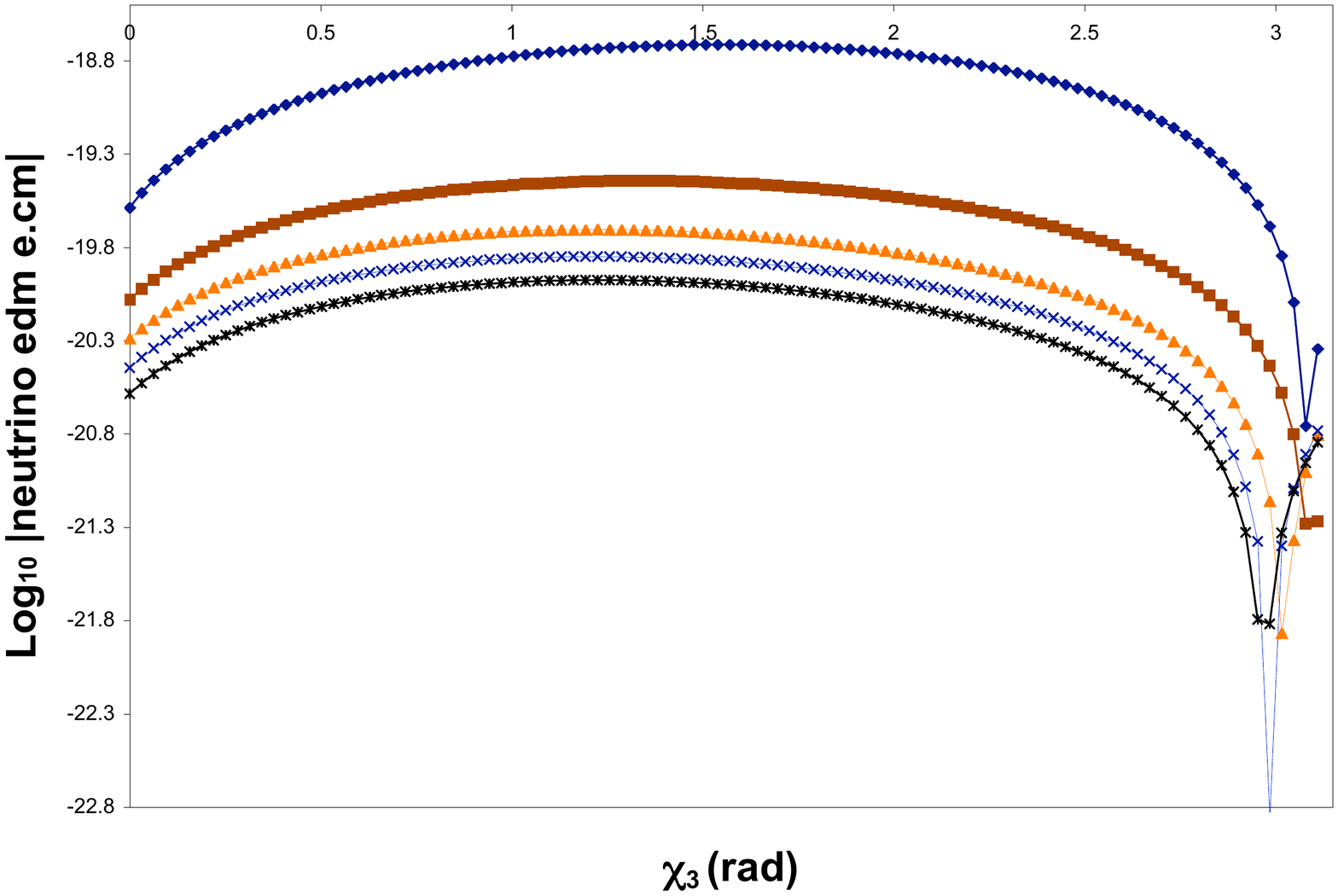}
 \includegraphics[width=10cm,height=12cm]{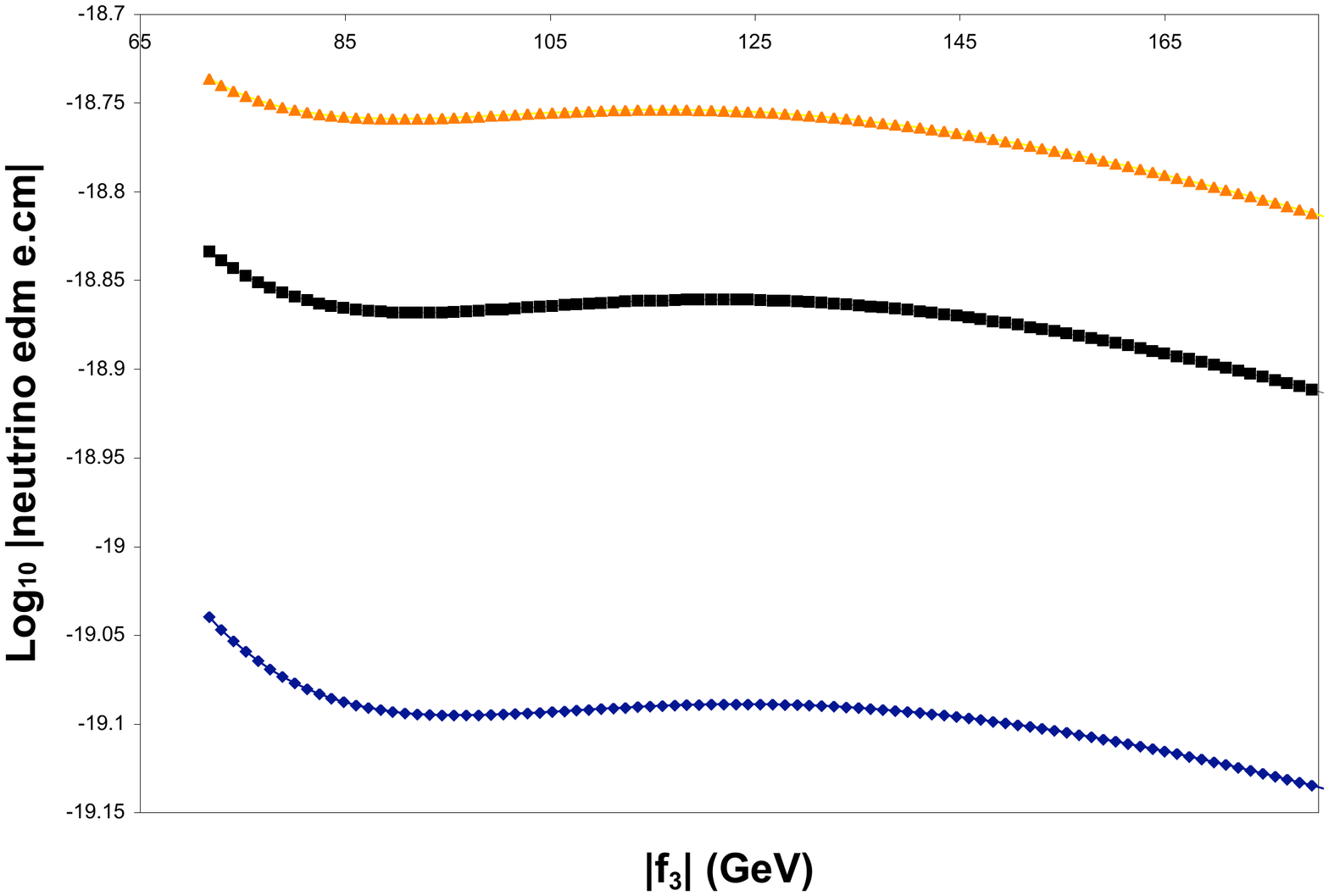}
    \includegraphics[width=10cm,height=12cm]{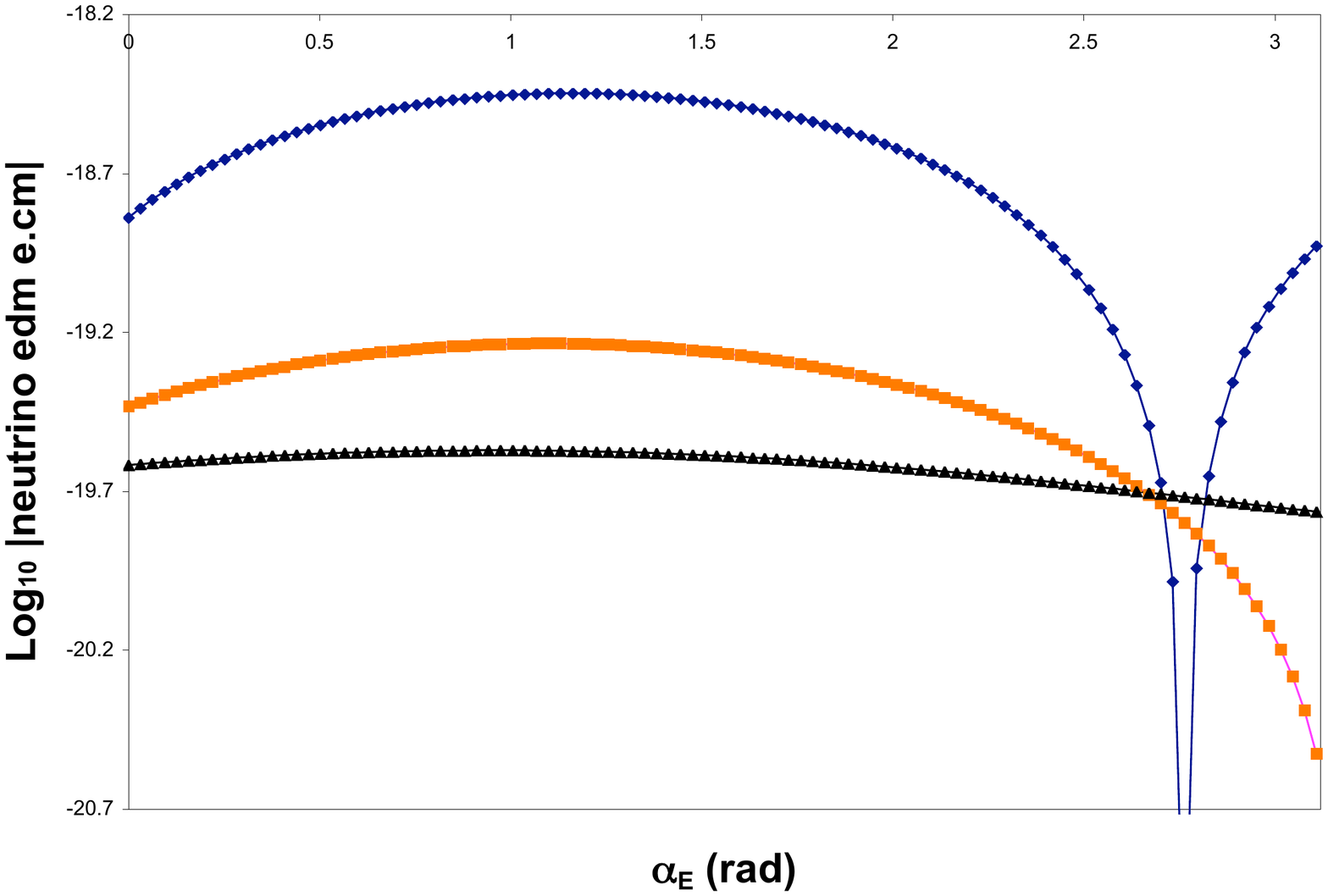}
      }     
\caption{\scriptsize Left:
An exhibition of the dependence of $d_{\nu}$ on $\chi_3$ with the input $\tan\beta=10$, 
 $m_E=120$, $|f_3|=$60, $|f_4|=$80,
$|f_5|=$100, $m_0=$100, $|A_0|=$170, $\tilde{m}_1=50$, $\tilde{m}_2=100$, $\mu=150$, $\chi_4=$0.2,
$\chi_5=$0.7, $\alpha_E=$0.6 and $\alpha_N=$0.4,  and 
$m_N=$300, 250, 200, 150, 100 (in ascending order at $\chi_3=0$).
Middle: 
An exhibition of the dependence of $d_{\nu}$ on $|f_3|$ with the input  $\tan\beta=10$,
$m_N=$120, $m_E=$100, $|f_4|=$80,
$|f_5|=$60, $m_0=$150, $|A_0|=$100, $\tilde{m}_1=50$, $\tilde{m}_2=100$, $\mu=150$ GeV and the phases $\chi_4=$0.3,
$\chi_5=$0.7, $\alpha_E=$0.4 and $\alpha_N=$1.0,  and $\chi_3=$0.4, 0.8, 1.2 (in ascending order).
Right:
An exhibition of the dependence of $d_{\nu}$ on $\alpha_E$ with the input $\tan\beta=20$
$m_E=120$, $|f_3|=$80, $|f_4|=$60,
$|f_5|=$90, $m_0=$100, $|A_0|=$120, $\tilde{m}_1=50$, $\tilde{m}_2=100$, $\mu=150$,
$\chi_3=$0.4, $\chi_4=$0.8, $\chi_5=$0.7 and $\alpha_N=$0.5, and 
$m_N=$190, 140, 90 (in ascending order at $\alpha_E=0$). 
 }
\label{fig4}
       \end{figure}

\end{document}